\documentclass[pra,groupedaddress,showpacs,twocolumn,nofootinbib]{revtex4-1}
\usepackage{amsmath,amssymb,amsthm}
\usepackage{graphicx}

\newtheorem{ThmEquiv}{Theorem}
\newtheorem{ThmEquivb}[ThmEquiv]{Theorem}
\newtheorem{ThmRoots}[ThmEquiv]{Theorem}
\newtheorem{ThmControl}[ThmEquiv]{Theorem}

\def\HC{{\cal H}}
\def\RC{{\cal R}}
\def\UC{{\cal U}}
\def\VC{{\cal V}}
\def\WC{{\cal W}}

\def\bra#1{\langle#1|}
\def\dyad#1#2{|#1\rangle\langle#2|}

\def\ket#1{|#1\rangle }
\newcommand{\ad}{^\dagger }
\newcommand{\ot}{\otimes }
\newcommand{\tP}{{\tilde P}}
\newcommand{\mat}[1]{\left(\begin{matrix}#1\end{matrix}\right)}

\long\def\ca#1\cb{}

 \def\outl#1{}  \def\xa{} \def\xb{}

\ca
 \def\outl#1{\par{\medskip\noindent\hspace*{.5cm}\bf
      \mathversion{bold}#1\mathversion{normal}\smallskip} }
 \long\def\xa#1\xb{}
 
\cb

\ca
 \def\outl#1{\par{\medskip\noindent\hspace*{.5cm}\bf
      \mathversion{bold}#1\mathversion{normal}\smallskip} }
 \def\xa{} \def\xb{}  
\cb

\begin{document}


\title{Fast protocols for local implementation of bipartite nonlocal unitaries}

\author{Li Yu$^1$}
\email{liy@andrew.cmu.edu}
\author{Robert B. Griffiths$^1$}
\author{Scott M. Cohen$^{1,2}$}
\affiliation{$^1$Department of Physics, Carnegie-Mellon University, Pittsburgh, Pennsylvania 15213, USA\\
$^2$Department of Physics, Duquesne University, Pittsburgh, Pennsylvania 15282, USA}

\begin{abstract}
  In certain cases the communication time required to deterministically
  implement a nonlocal bipartite unitary using prior entanglement and LOCC
  (local operations and classical communication) can be reduced by a factor of
  two.  We introduce two such ``fast'' protocols and illustrate them with various
  examples. For some simple unitaries, the entanglement resource is used quite efficiently.
  The problem of exactly which unitaries can be implemented by these two protocols
  remains unsolved, though there is some evidence that the set of implementable unitaries
  may expand at the cost of using more entanglement.
\end{abstract}

\pacs{03.67.Ac, 03.67.Dd, 03.67.Lx}
\maketitle



\section{Introduction}\label{sct1}

A central problem in quantum information theory is that of interconversion
between resources, for example, how to use communication channels to produce
entanglement between distant parties, or how to use such entanglement to carry
out nonlocal operations.  In particular, the use of prior entanglement
assisted by classical communication to carry out nonlocal unitaries has been
the subject of various studies \cite{Eisert,ReznikNLU,NLU}; for a more
extensive list see Ref.~\cite{NLU}.

In this paper we add \emph{time} as a resource to be considered along with entanglement cost when constructing
protocols for bipartite nonlocal unitaries (nonlocal gates). The ability to implement nonlocal unitaries rapidly may be particularly relevant in the context of distributed quantum computation \cite{Cirac_dist,Yims,Yims2,Meter}, where less time consumption means less decoherence; or in position-based quantum cryptography \cite{Kent,Kent06,Lau,Buhrman}, where it may allow certain position verification schemes to be broken.

The usual protocols for bipartite unitaries, such as those in Ref.~\cite{NLU}, have the following
general structure: Alice carries out local operations and measurements, and
sends the measurement results through a classical communication channel to Bob, who then carries out corresponding operations and
measurements, and sends the measurement results back to Alice using classical
communication. Finally, Alice performs additional local operations that may depend on the previous measurement results of both parties.
When the distance between the two parties is large the total time required for the protocol will be dominated by
the two rounds of communication, thus double the minimum time for a signal to
pass from one to the other.  However, there exist nonlocal unitaries
which can be implemented by a protocol in which Alice and Bob carry out local operations and measurements at the same time,
and then simultaneously send the results to the other party, and finally perform local operations depending on the received messages to complete the protocol. This reduces the total communication time by a factor of two \footnote[1]{There may be in practice other temporal costs that need to be taken into account, such as that required to produce the initial entangled state. We are ignoring these in the present paper.}.  We are interested in identifying which
bipartite unitaries can be carried out using such a \emph{fast} protocol, and
also in finding the associated entanglement cost.  The crucial distinction between a
fast and slow protocol of the form considered here is that for the latter, Bob
needs to wait for a message from Alice before choosing the basis in which to carry
out his measurement (``choosing the measurement basis'' is equivalent to choosing what local gates to do before his measurement),
whereas in the former this basis can be fixed in advance.

We have identified two classes of nonlocal unitary that lend themselves to a
fast protocol: \emph{controlled} unitaries of the form shown in \eqref{eqn1}
below, and \emph{group} unitaries of the form shown in \eqref{eqn9}.  The slow
versions of both were considered in our previous work \cite{NLU}, where we
showed that controlled unitary protocols, while useful for understanding how
such protocols work, can always be replaced by group unitary protocols that
make use of the same resources.  Our fast protocols represent special cases
(i.e., special groups and parameter choices) of the slow protocols discussed
previously, and once again the controlled kind can be replaced by the group
kind.  By increasing the amount of entanglement expended, additional unitaries
can be carried out using these fast protocols.  In some cases this
allows an arbitrarily close approximation to a unitary which cannot
be carried out exactly by these methods.
A still more general class of slow protocols, corresponding to Eq.~(18)
in \cite{NLU}, also has a fast version, but we have yet to find examples of
unitaries it can carry out that cannot be implemented by our other fast
protocols.

The protocols we consider are deterministic---they succeed with probability
one---and use a definite amount of entanglement determined in advance.  Such
deterministic fast protocols have previously been studied by Groisman and
Reznik \cite{Groisman05} for a controlled-NOT (CNOT) gate on two qubits, and by Dang and Fan
\cite{Dang} its counterpart on two qudits.
In addition, Buhrman {\it et al.} \cite{Buhrman} and Beigi and K\"onig
\cite{Beigi11} have published approximate schemes for what they call
``instantaneous quantum computation,'' equivalent to a fast bipartite unitary
in our language.  These protocols, unlike ours, can be used, to approximately
carry out any bipartite unitary.  The one in \cite{Buhrman},
which is based on the nonlocal measurement protocol in \cite{Vaidman}, has a probability
of success less than 1, so it is not deterministic, but this probability can
be made arbitrarily close to 1 by using sufficient entanglement.  The protocol
in \cite{Beigi11} uses a fixed amount of entanglement to implement with
probability 1 a bipartite quantum operation (completely positive trace
preserving map) which is close to the desired unitary, and it can be made
arbitrarily close by using sufficient entanglement.  The term
``instantaneous'' is not unrelated to the idea of an ``instantaneous
measurement'' as discussed in \cite{Vaidman,Groisman02,Clark10}, where the
terminology seems somewhat misleading in that completing their protocols
actually requires a finite communication time, e.g., the parties must send the
results to headquarters (or to each other) in order to complete the
identification of the measured state.  In the same way, ``instantaneous
quantum computation'' actually requires a finite communication time, the same
as in our fast protocol.

The paper is organized as follows.  In Sec.~\ref{sct2} we consider
controlled unitaries of the form Eq.~\eqref{eqn1}, where the unitaries being
controlled form an Abelian group.  (Appendix~\ref{sct_extend} contains an
argument, which may be of more general interest, that allows projectors $P_k$
in this formula to be replaced by projectors of rank 1.)
In addition we show how subsets of the collection of unitaries representing an Abelian group
can be employed to generate fast unitaries otherwise not accessible by
our protocol. Section~\ref{sct3} is devoted to group unitaries of the form
\eqref{eqn9}, including a significant number of examples.  We also present
an argument showing that the controlled-Abelian-group unitaries of
Sec.~\ref{sct2} can be transformed to group unitaries of the form
\eqref{eqn9}.  The concluding Sec.~\ref{sct4} contains a brief summary
along with an indication of some open problems.

\section{Fast protocol for controlled-Abelian-group unitaries}\label{sct2}

In this section we construct a fast protocol for any
controlled-Abelian-group unitary of the form
\begin{equation}\label{eqn1}
\UC=\sum_{k=0}^{N-1} P_k \otimes V_k,
\end{equation}
where the $P_k$ are orthogonal projectors, possibly of rank greater than 1, on
a Hilbert space $\HC_A$ of dimension $d_A$.  The $\{V_k\}$ are unitary
operators on a Hilbert space $\HC_B$ of dimension $d_B$, that form a
representation of an Abelian group $G$ of order $N$. As shown in
Appendix~\ref{sbct_control_rank}, it suffices to consider projectors of rank 1.
That is, a scheme for implementing
\begin{equation}\label{eqn2}
\UC=\sum_{k=0}^{N-1} \dyad{k}{k} \otimes V_k,
\end{equation}
where $\ket{k}$ denotes a ket belonging to a standard (or computational)
orthonormal basis, is easily extended to one that carries out the more general
\eqref{eqn1}.  In addition we shall consider cases,
Sec.~\ref{sbct2.3}, in which the $\{V_k\}$ form a \emph{subset}
of an Abelian group, with the sum in \eqref{eqn1} restricted to a subset $S$ of
the integers from $0$ to $N-1$.

\subsection{Fast protocol for controlled-cyclic-group unitaries}\label{sbct2.1}

\begin{figure*}[ht]
\begin{centering}
\includegraphics[scale=1]{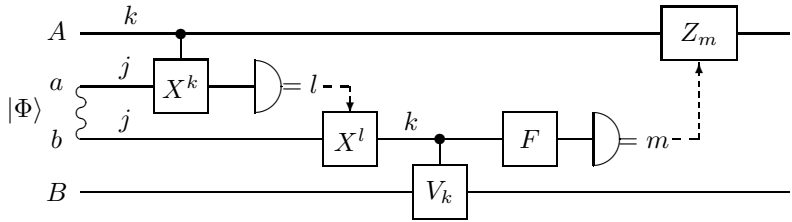}
\caption{The slow protocol in Sec.~III of \cite{NLU} for implementing the
unitary $\UC=\sum_{k=0}^{N-1} \dyad{k}{k} \otimes V_k$.}
\label{fgr1}
\end{centering}
\end{figure*}

The simplest Abelian group is a cyclic group, so we start with the case where
the $\{V_k\}$ in \eqref{eqn2} are a representation of such a group. (It
suffices to consider ordinary representations, since a projective
representation of a cyclic group is equivalent to an ordinary representation;
see Sec.~12.2.4 of \cite{GroupBook}.)  It will be convenient to let $V_0$ be the
identity and $V_k = V_1^k$. The slow protocol \cite{NLU} for this case, which
works for any collection $\{V_k\}$ of unitaries on $B$, is shown in
Fig.~\ref{fgr1}, where
\begin{equation}\label{eqn3}
\ket{\Phi} = \frac{1}{\sqrt{N}}\sum_{j=0}^{N-1} \ket{j}_a\otimes \ket{j}_b
\end{equation}
is a fully entangled state on the ancillary systems $a$ and $b$ associated
with $A$ and $B$, respectively, and
the gates $X$, $Z$, and $F$ (the Fourier
matrix) are defined by
\begin{align}\label{eqn4}
X&=\sum_{j=0}^{N-1}\dyad{j\ominus 1}{j},\quad
Z=\sum_{k=0}^{N-1} e^{2\pi ik/N}\dyad{k}{k},\notag\\
F&=\frac{1}{\sqrt{N}}\sum_{m,k=0}^{N-1}e^{2\pi i m k/N}\dyad{m}{k}.
\end{align}
Here $j\ominus 1$ denotes $(j-1)\mod N$,
so $X^l\ket{j}=\ket{j\ominus l}$.  The symbols resembling ``D'' in Fig.~\ref{fgr1}
represent measurements in the standard basis.

The slow protocol proceeds by Alice carrying out the operations indicated on
the left side of Fig.~\ref{fgr1}, and then sending the outcome $l$ of
the measurement on $a$ to Bob over a classical channel. He uses it to carry
out a gate $X^l$, followed by the other operations in the center of the
figure.  His final measurement outcome $m$ is sent to Alice over another
classical channel, who uses it to perform an additional gate $Z_m = Z^{-m}$
that completes the protocol.

\begin{figure*}[ht]
\begin{centering}
\includegraphics[scale=1]{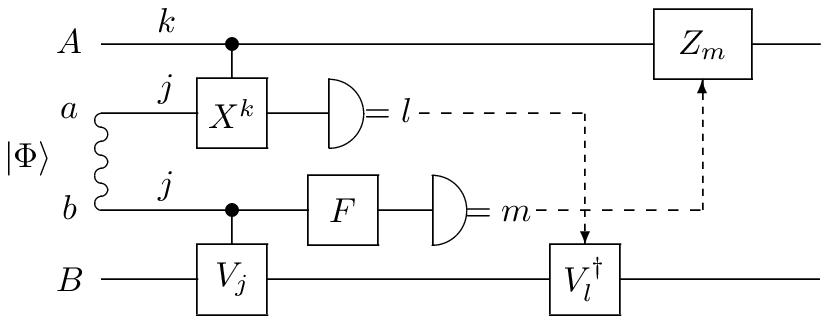}
\caption{The fast protocol for implementing the unitary
$\UC=\sum_{k=0}^{N-1} \dyad{k}{k} \otimes V_k$.}
\label{fgr2}
\end{centering}
\end{figure*}

A faster protocol can be constructed if the two rounds of classical
communication can be carried out simultaneously instead of consecutively.
This is possible if Bob can carry out various operations, including a
measurement, in advance of receiving the value of $l$ from Alice, as in
Fig.~\ref{fgr2}.  The classical signals can then be sent simultaneously,
and the protocol is completed when both Alice and Bob make final corrections
that depend on the signals they receive.  In order to change the slow protocol
into a fast protocol, one must, in effect, push the $X^l$ gate in
Fig.~\ref{fgr1} through the two gates that follow it in order to arrive
at the situation in Fig.~\ref{fgr2}. The two steps are as follows:
\begin{enumerate}
  \item Commute $X^l$ with the controlled-$V_k$ gate: the $X^l$ itself
passes through the control node unchanged, but leaves the $V$ gate controlled
by state $|j\rangle$ instead of $|k\rangle$, so $V_j$ instead of $V_k$
acts on $B$.  This can be compensated at the end of the
protocol by a local unitary correction of $V_l^\dag$, with $l=j\ominus k$ (see the discussion below).
\item Commute $X^l$ with $F$: we have that $F X^l = Z^{-l} F$, and since the
  $Z^{-l}$ are diagonal unitaries, they do not affect the measurement result
  $m$ in the standard basis, and thus $Z^{-l}$ is absent from the fast
  protocol in Fig.~\ref{fgr2}. (Due to the removal of $Z^{-l}$ there is an
  unimportant global phase, dependent on $l$ and $m$, that is introduced in
  the implementation of $\UC$.)
\end{enumerate}

The final correction $V_l\ad$ is possible because the $\{V_k\}$ form a cyclic
group: The net operation on $B$ is $V_l^\dag
V_j=V_{j\ominus l}=V_{j\ominus (j\ominus k)}=V_k$, where $l=j\ominus k$
follows from Fig.~\ref{fgr1} and the definition of the $X^l$ gate.
This is an extra restriction over the slow protocol, where the $V_k$ can be
arbitrary unitaries.\smallskip

\textbf{Example 1. }

\noindent
Case (i). $\UC$ is the $N$-dimensional CNOT
gate, with $N=d=d_A=d_B$, and $V_k=X^k$ form a cyclic group. This is the class
of unitaries implementable by Dang and Fan's protocol shown in Fig.~2 of
\cite{Dang}.\smallskip

\noindent
Case (ii). $\UC$ is a controlled unitary of the form \eqref{eqn2} with
$N=d_A=3$, $d_B=2$, $V_k=\mbox{diag}(1,e^{2\pi i k/3})$, $k=0,1,2$. (The $V_k$
are not shift operators, so this is more general than the protocol in
\cite{Dang}.)

\subsection{Generalization to a controlled-Abelian-group}\label{sbct2.2}

The fast protocol for controlled-cyclic-group unitaries is easily generalized
to the case where the $V_k$ in \eqref{eqn1} form an ordinary representation of an Abelian group $G$ of order $N$.  Again it suffices
(Appendix~\ref{sbct_control_rank}) to consider the case \eqref{eqn2} where the
$P_k$ are rank 1 projectors. Any finite Abelian group is the direct sum
(direct product) of $\eta$ cycles, and it is convenient to adopt a label $k$
for elements of $G$
that reflects this structure, by thinking of it as an $\eta$-tuple of integers,
\begin{equation}\label{eqn5}
k=(k_1,k_2,\ldots k_{\eta}),
\end{equation}
with $0\le k_i\le r_i-1$, where $r_i$ is the length of the $i$-th cycle.
In this way group multiplication, with $k=(0,0,\ldots 0)$ the identity, is the
same as vector addition, modulo $r_i$ for the $i$-th component. Similarly, the $j$
labels on the systems $a$ and $b$ in Fig.~\ref{fgr2}, and the measurement outcomes $l$ and $m$,
can also written as $\eta$-tuples: $j=(j_1,j_2,\ldots j_{\eta})$, etc. In the following
we will make use of the inner product of two $\eta$-tuples such as $(j\cdot m) = \sum_{i=1}^{\eta} j_i m_i$.

The $X$, $Z$, and $F$ gates are now appropriate tensor products of the cyclic
group gates in \eqref{eqn4}, for example, $X^k$ understood as
$\bigotimes_{i=1}^{\eta} X_i^{k_i}$ and $X^k\ket{j}=\ket{j\ominus k}$, using
the obvious $\eta$-tuple definition of $j\ominus k$.  The $Z_m$ gate
in Fig.~\ref{fgr2} is the tensor product of the $Z_i^{-m_i}$ gates for the
different cycles,
\begin{equation}
\label{eqn6}
Z_m=\sum_k e^{-2\pi i (k \cdot m)/N} \dyad{k}{k}.
\end{equation}

Here is why the protocol works. Assume an initial product
state $\ket{k}\ot\ket{\omega_k}$ on $\HC_A\ot\HC_B$.
Then the operator implemented on $B$ is $V_l^\dag V_j=V_{j\ominus k}^\dag
V_j=V_{k\ominus j} V_j=V_k$. The $F$ gate on $b$ before the measurement gives
rise to a phase $e^{2\pi i (j \cdot m)/N}$, which is partially compensated by the
phase $e^{-2\pi i (k \cdot m)/N}$ in the $Z_m$ gate on $A$,
and since $j=k\oplus l$, we are left with an overall phase of $e^{2\pi i (l \cdot m)/N}$.
Since this phase is independent of $k$, a superposition of initial product states of this form for
different $k$ will also be transformed by $\UC$, up to an overall phase that
is of no concern.

Note that the $V_k$ themselves may, but need not, be tensor
products, as illustrated in the following example.\smallskip

\textbf{Example 2.}

\noindent
Case (i): $d_A = d_B=4$. The $V_k$ defined by
\begin{align}\label{eqn7}
& V_{(0,0)}=\mbox{diag}(1,1,1,1),\, & V_{(0,1)} & =\mbox{diag}(1,1,-1,-1),
\notag\\
& V_{(1,0)}=\mbox{diag}(1,-1,1,-1),\, & V_{(1,1)} & =\mbox{diag}(1,-1,-1,1),
\end{align}
are tensor products, and form a group $C_2\times C_2$.  If one regards
$\HC_A$ as well as $\HC_B$ as a tensor product of two qubits, the $\UC$
defined by \eqref{eqn2} is itself a tensor product $\UC = \UC_1 \otimes
\UC_2$, with each factor a controlled-cyclic group unitary with one qubit on
the $A$ and the other on the $B$ side.  Thus $\UC$ can be implemented by an
overall protocol which is just two smaller protocols running in parallel with
each other, one for $\UC_1$ and the other for $\UC_2$.\smallskip

\noindent
Case (ii): $d_B=3$.  Modify the $V_k$ in \eqref{eqn7} by keeping only the
first three rows and columns, so they are no longer tensor products, though
they still form a group $C_2\times C_2$.  Consequently, the protocol that
carries out $\UC$ can no longer be viewed as two smaller protocols running in
parallel.

\subsection{Controlled subset of an Abelian group}
\label{sbct2.3}

Assume that the $\{V_k\}$ in \eqref{eqn1} form an ordinary representation of
an Abelian group of order $N$, but the sum over $k$ is restricted to some
subset $S$ of the set of $N$ $\eta$-tuples defined in the last subsection.  It
will suffice once again to consider the case of rank-one projectors, i.e.,
\eqref{eqn2}.  The $j$, $l$, and $m$ in Fig.~\ref{fgr2} run over the same
range as before, but $k$ is restricted to the set $S$.  Therefore the
dimension $d_A=n$ of $\HC_A$ is less than the Schmidt rank of $\ket{\Phi}$,
which is the order $N$ of the group.  It is convenient to use the elements of
$S$ to label the kets that form the basis of $\HC_A$ in \eqref{eqn2}
[corresponding to the projectors in \eqref{eqn1}]. The operator $Z_m$ is now
given by \eqref{eqn6}, but with $k$ restricted to $S$.
The reason that the fast
protocol in Fig.~\ref{fgr2} will work in this case is the same as given above
in Sec.~\ref{sbct2.2}; the fact that $k$ is restricted to a subset makes
no difference.

The significance of this extension of the result in Sec.~\ref{sbct2.2} is that
it enlarges the class of fast unitaries that can be carried out using a protocol
of this sort, though perhaps with a significant increase in the entanglement
cost.  This is illustrated by the following example, which shows that in
certain cases one can approximate a continuous family of unitaries using sufficient entanglement (a large enough $N$).\smallskip

\textbf{Example 3.}

Consider a unitary on two qubits $A$ and $B$ of the form
\begin{equation}\label{eqn8}
\UC=\dyad{0}{0}\otimes I +\dyad{1}{1}\otimes R,
\end{equation}
where $R = V_m = \mbox{diag}(1,e^{2\pi i m/N})$ for some integer $m < N$. By
relabeling $\dyad{1}{1}$ on $A$ as $\dyad{m}{m}$, we see that \eqref{eqn8}
is of the form \eqref{eqn2} with $k$ in the sum restricted to the two values
in $S=\{0,m\}$. Thus $\UC$ can be carried out in a fast way at an
entanglement cost of $\log_2 N$ ebits. In general, any two-qubit controlled unitary is of the form \eqref{eqn8} (up to local unitaries on $A$ and $B$, before and after $\UC$) with $R =\mbox{diag}(1,e^{i\phi})$ for some real number $\phi$. Since $\phi$ can be approximated by $2\pi$ multiplied by a rational number $m/N$ with large enough $N$, any two-qubit controlled unitary can be approximately implemented (up to local unitaries) using this fast protocol by setting $R$ equal to $\mbox{diag}(1,e^{2\pi i m/N})$ for suitable $m$ and $N$; the entanglement cost is again $\log_2 N$ ebits.

A further generalization to arbitrary $d_A$ and $d_B$ is possible for any unitary $\UC$ which is diagonal in a product of bases, a basis on $\HC_A$ and another on $\HC_B$. When such a diagonal $\UC$ is written in the form \eqref{eqn2}, the unitaries $V_k$ are diagonal. Each diagonal element of $V_k$ is approximately an integer root of unity, hence each $V_k$ is approximately an integer root of the identity operator, and the whole set $\{V_k\}$ can be approximated by a subset of an ordinary representation of an Abelian group of sufficient size. Thus any bipartite unitary diagonal in a product of bases can be approximately implemented by a fast protocol.

\section{Fast protocol for double-group unitaries}
\label{sct3}

In this section we consider a fast protocol for ``double-group'' unitaries
of the form
\begin{equation}\label{eqn9}
    {\cal U}=\sum_{f\in G}c(f)\,\Gamma(f)
=\sum_{f\in G}c(f)\,U(f)\otimes V(f),
\end{equation}
where $G$ is a group of order $N$, $U(f)$ and $V(f)$ are unitaries on Hilbert
spaces $\HC_A$ and $\HC_B$ of dimension $d_A$ and $d_B$, respectively, and
the operators $\Gamma(f) = U(f)\ot V(f)$ form a projective representation of
$G$, in the sense that
\begin{equation}
 \Gamma(g) \Gamma(h) = \lambda(g,h) \Gamma(gh).
\label{eqn10}
\end{equation}
The collection $\{\lambda(g,h)\}$ of complex numbers of unit magnitude
is known as the factor system.  Similarly, $\{U(f)\}$ and $\{V(f)\}$ each
form a projective unitary representation of $G$ with individual factor systems
which may differ from one another, whose product for a given $g$ and $h$ is
$\lambda(g,h)$. From Sec.~12.2.1 of \cite{GroupBook}, for our purposes we can assume
the factor system $\{\lambda(g,h)\}$ is \emph{standard}, that is,
\begin{equation}
\lambda(e,e)= \lambda(e,f)= \lambda(f,e)=1,\,\,\,\forall f\in G,
\label{eqn11}
\end{equation}
where $e$ is the identity element in $G$.

\subsection{Protocol}\label{sbct3.1}

\begin{figure*}[ht]
\begin{centering}
\includegraphics[scale=1]{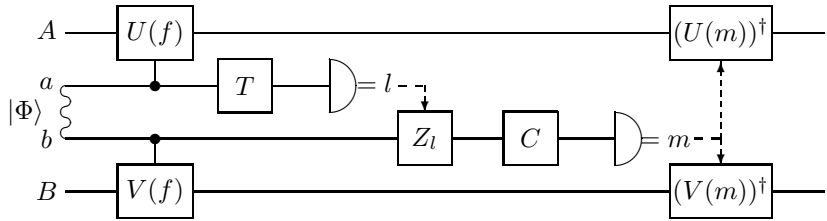}
\caption{The slow protocol that implements the unitary ${\cal U}$
in \eqref{eqn9}.} \label{fgr3}
\end{centering}
\end{figure*}

The slow protocol in Sec.~IV~D of \cite{NLU} for implementing unitaries of the
type \eqref{eqn9} is shown in Fig.~\ref{fgr3}. The two parties share a
maximally entangled state
\begin{equation}\label{eqn12}
\ket{\Phi} = \frac{1}{\sqrt{N}}\sum_{f\in G} \ket{f}_a\otimes \ket{f}_b,
\end{equation}
on the ancillary systems $a$, $b$, each of dimension $N$, the order of $G$.
Alice and Bob perform controlled-$U(f)$ and controlled-$V(f)$ gates on $aA$
and $bB$, respectively. Alice follows this with a $T$ gate on $a$, where in
the standard basis $T$ is a complex Hadamard matrix $\hat T$ divided by
$\sqrt{N}$, that is, a unitary with all elements of the same magnitude
$1/\sqrt{N}$. Then she does a measurement on $a$ in the standard basis and
sends the result $l$ to Bob. Bob carries out a $Z_l$ gate on $b$, where each
$Z_l$ is a diagonal unitary matrix whose diagonal elements are the complex
conjugates of those in the $l$-th row of $\hat T$.  Thus $T$ and $Z_l$
generalize the Fourier gate $F$ and the $Z^{-l}$ gate in our previous paper
\cite{NLU}. This more general choice does not extend the set of unitaries the
slow protocol can implement, since the phases in $T$ and $Z_l$ cancel each
other, but it allows the fast protocol to implement a larger set of unitaries
than would otherwise be possible.

Next, Bob applies a unitary gate
\begin{equation}
\label{eqn13}
C = \sum_{f,g\in G} \lambda(g,g^{-1}f)\, c(g^{-1}f)\; \dyad{g}{f}
\end{equation}
to $b$, where $\lambda(g,h)$ is defined in \eqref{eqn10} and the coefficients
$c(f)$ are those in \eqref{eqn9}. (The coefficients $c(f)$ are not uniquely
defined by $\UC$ if the $\Gamma(f)$ are linearly dependent, but there is
always at least one choice for which $C$ is unitary, Theorem~7 of \cite{NLU}.)
Then he measures $b$ in the standard basis and sends the result $m$ to Alice.
To complete the protocol Alice and Bob apply unitary corrections $(U(m))^\dag$
and $(V(m))^\dag$ to their respective systems.

\begin{figure*}[ht]
\begin{centering}
\includegraphics[scale=1]{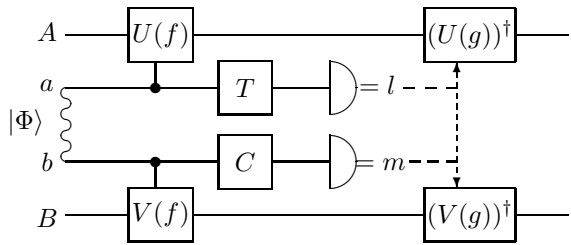}
\caption{The  fast protocol that implements the unitary $\UC$ in \eqref{eqn9}
when the conditions given in Theorem~\ref{thm1} are satisfied. The $g$ for the
final corrections depends on both measurement outcomes $l$ and $m$. }
 \label{fgr4}
\end{centering}
\end{figure*}

\outl{Introducing the fast protocol by modifying the slow protocol.}

The slow protocol in Fig.~\ref{fgr3} can be replaced with the fast protocol
in Fig.~\ref{fgr4} provided the $Z_l$ gate in the former,  which in effect
determines the basis for Bob's measurement, can be eliminated at the cost of
re-interpreting the outcome $m$ of his measurement.  A sufficient condition
for this is that for every $l$ there exists a \emph{complex permutation}
matrix $\tP_l$ (exactly one nonzero element of magnitude 1 in each row and
in each column) such that
\begin{equation}
\label{eqn14}
C Z_l= {\tilde P}_l C.
\end{equation}
The effect of $\tP_l$ is simply to permute the measurement outcomes, which can
then be re-interpreted once $l$ is known. The phases in $\tP_l$ would only introduce a
global phase for the implemented $\UC$ (dependent on $l$ and $m$) and are therefore of no concern.

A useful procedure for generating $C$ and $T$ matrices for which \eqref{eqn14}
holds employs the \emph{character table} $\hat K$ of an Abelian group $H$ of
order $N$.  This is an $N\times N$ matrix, all elements of which are of
magnitude 1, with columns labeled by elements of $H$ and rows by its distinct
irreducible representations, all of which are one-dimensional.  Because the
representations are one-dimensional, each row is itself a representation
(i.e., the character is the representation ``matrix''). The
element-wise product of two columns is another column, since each row is a
representation of the group; likewise the element wise product of two rows is
another row, since the (tensor) product of two representations is a
representation.  Thus the transpose of a character table is again a character
table.  The complex conjugate of any column (or row) is another column (row)
of $\hat K$ corresponding to the inverse of the group element. The actual
order of the columns or the rows is arbitrary, though it is often convenient
to assume that the first row and the first column contain only $1$'s.  Since
$K=\hat K/\sqrt{N}$ is a unitary matrix, a character table $\hat K$ is a
special case of a complex Hadamard matrix: one whose elements are all of
magnitude 1, and whose rows (and columns) are mutually orthogonal.  If $H$ is
a cyclic group $C_N$, then up to permutations of rows or columns $\hat K =
\hat F = \sqrt{N}\, F$, with $F$ the Fourier matrix \eqref{eqn4}; similarly if
$H$ is the direct product (sum) of cycles, $\hat K$ is the tensor product of
the corresponding Fourier matrices \cite{GroupProduct}.

\begin{ThmEquiv}
\label{thm1}
\

(a) Let $\hat K$ be the character table of an Abelian group $H$ of order $N$,
and define

\begin{equation}
\label{eqn15}
 \hat T = \sqrt{N}\, T = L\hat K,\quad
  \hat C = \sqrt{N}\, C = M\hat K D,
\end{equation}
where $L$ and $M$ are complex permutation matrices, and $D$ a diagonal matrix
with diagonal elements of magnitude 1, thus a diagonal unitary.
Let $Z_l$ be the diagonal matrix with diagonal elements equal to the complex
conjugates of those forming row $l$ of $\hat T$. Then there exist
complex permutation matrices $\tilde P_l$ such that \eqref{eqn14} is satisfied
for every $l$.

(b) If the rows of an $N\times N$ matrix $\hat T$ are linearly independent and
form a (necessarily Abelian) group $H$ up to phases under element-wise multiplication, then $\hat T$ is of the form given in \eqref{eqn15}.
\end{ThmEquiv}

The proof is in Appendix~\ref{appthm1}. Note that the group $H$ need not be
isomorphic to the group $G$ represented by $\{\Gamma(f)\}$ although they are
of the same order---see Example 8 in Sec.~\ref{sbct3.3} with $n>2$.  The
matrix $\hat T$ defined in \eqref{eqn15} has the property that the rows under
element-wise products form the Abelian group $H$ up to a possible phase factor
determined by $L$.

One consequence of \eqref{eqn15} is
\begin{equation}
\label{eqn16}
 C= PT D,
\end{equation}
where $P$ is a complex permutation matrix and $D$ is a diagonal unitary.  When
the matrix $C$ is of the form \eqref{eqn13}, the fact that all its elements
are of the same magnitude $1/\sqrt{N}$, as implied by \eqref{eqn15}, means
that the same is true of the $c(f)$.  A partial answer to the question of
whether \eqref{eqn14} implies that $T$ and $C$ have the form given in
\eqref{eqn15} is provided by the following theorem, whose proof is also in
Appendix~\ref{appthm1}.

\begin{ThmEquivb}
\label{thm2}

For each $l$ let $Z_l$ be the diagonal matrix whose diagonal elements are the complex
conjugates of those forming row $l$ of a complex Hadamard matrix $\hat T
=\sqrt{N}\, T$.  If there exists a unitary matrix $C$ without a zero element
in its first row, and complex permutation matrices ${\tilde P}_l$ such
that Eq.~\eqref{eqn14} holds for every $l$, then the matrices $Z_l$ form an Abelian
group up to phases, and \eqref{eqn15} and \eqref{eqn16} hold.

\end{ThmEquivb}

It is worth noting that with $T$  and $C$ of the form \eqref{eqn15} there is a
symmetrical version of  the fast protocol in Fig.~\ref{fgr4}  in which the $C$
gate on the  $B$ side is replaced with $T$.  This  requires that the entangled
resource $\ket{\Phi}$ be changed to
\begin{equation}\label{eqn17}
\ket{\Phi'} = (I\otimes D)\ket{\Phi}.
\end{equation}
The reason this works is that the changes produced by $P$ in \eqref{eqn16} in
the measurement outcome $m$ can always be compensated by altering the function
$g(l,m)$ that determines the final corrections.  The following is a simple
example.\smallskip

{\bf Example 4.}

The two-qubit unitary,
\begin{equation}\label{eqn18}
\UC=\frac{1}{\sqrt{2}} (I_A\otimes I_B + i Z_A\otimes Z_B),
\end{equation}
where $Z_A$ and $Z_B$ are Pauli $\sigma_z$  gates on
$A$ and $B$, is equivalent under local unitaries to a CNOT gate.  It is of
the form \eqref{eqn9} with $G$ the cyclic group $C_2$ of order 2, and can be
implemented by the fast protocol in Fig.~\ref{fgr4} using the matrices
\begin{equation}\label{eqn19}
T = \frac{1}{\sqrt{2}}\mat{
1 & 1 \\
1 & -1}, \quad
C=\frac{1}{\sqrt{2}} \mat{
1 & i \\
i & 1},
\end{equation}
where the rows of $T$ multiplied by $\sqrt{2}$ form a group $H=C_2$.
Because $T$ and $C$ change the measurement basis,
Alice and Bob effectively perform measurements of $\sigma_x$ and $\sigma_y$,
respectively; this is the same thing (with the parties interchanged) as the
fast protocol in \cite{Groisman05}. An equivalent symmetrized protocol in
which $C$ is replaced by $T$ employs a resource state
$\ket{\Phi'}=\frac{1}{\sqrt{2}}(\ket{00}+i\ket{11})$, and both parties
perform a $\sigma_x$ measurement.

\subsection{Which unitaries can be carried out using the fast protocol?}
\label{sbct3.2}

Given a particular bipartite unitary $\UC$, can it be implemented using the
fast double-group protocol?  Any such $\UC$ can always be written in the form
\eqref{eqn9} using a sufficiently large group (see Sec.~V A of \cite{NLU}),
and typically there are many different ways of constructing such an expansion.
However, for our fast protocol to work, assuming that $T$ and $C$ are of the
form given in \eqref{eqn15}, one must find a \emph{particular} expansion, a
particular group $G$ and unitaries $U(f)$ and $V(f)$ along with expansion
coefficients $c(f)$, that satisfy appropriate conditions.  In particular, (i)
the $c(f)$ must all be of the same magnitude $1/\sqrt{N}$, as noted following
\eqref{eqn16}.  But two additional conditions must be checked: (ii) the matrix
$C$ defined in \eqref{eqn13} must be unitary, and (iii) $C$ must be related to
the character table of some group $H$ as in \eqref{eqn15}.  Condition (iii)
can be checked in the following way. Multiply each row and then each column of
$\hat C=\sqrt{N}\,C$ by a suitable phase such that the resulting matrix $\hat
C' = Q\hat CR$, where $Q$ and $R$ are diagonal unitaries, has $1$'s in the
first row and first column.  Then check whether its rows (alternatively, its
columns) form a group $H$ under component-wise multiplication.  If this is so,
then $\hat C'$ is a character table of $H$. Equating it to $\hat K$ in
\eqref{eqn15}, letting $D=R^{-1}$ and $M = Q^{-1}$, and choosing any complex
permutation matrix $L$, we arrive at a $T$ which along with this $C$ satisfies
the conditions of Theorem~\ref{thm1}.  Thus, provided (i), (ii), and (iii) are
satisfied there is in fact a fast unitary $\UC$.

The scheme just described provides a useful approach
for constructing examples.  Start with a group $G$ and (projective) unitary
representations $\{U(f)\}$ and $\{V(f)\}$ on $\HC_A$ and $\HC_B$, and look for
a set of coefficients $\{c(f)\}$ of equal magnitude such that $C$ given by
\eqref{eqn13} is unitary, and satisfies condition (iii) in the preceding
paragraph.  The search is aided by noting that any factor system, see
Sec.~12.2.2 of \cite{GroupBook}, is equivalent to a \emph{normalized} factor
system in which each $\lambda(g,f)$ is an $N$-th root of 1.  A consequence,
proved in Appendix~\ref{appthm1}, is the following:

\begin{ThmRoots}
\label{thm3}
Let $\Gamma(f)$ in \eqref{eqn9} be a projective representation of a group $G$
of order $N$ with a normalized (see above) factor system $\lambda(g,h)$,
\eqref{eqn10}. Assume the matrix $C$ defined in \eqref{eqn13} is of the form
given in \eqref{eqn15}.  Then the coefficients $\{c(f)\}$ in \eqref{eqn13}
can be written in the form
\begin{equation}
  c(f)=(\gamma/\sqrt{N}\,) \exp[\,2\pi i k(f)/N^2\,],
\label{eqn20}
\end{equation}
where $k(f)$ is an integer that depends upon $f$, and $\gamma$ is a phase
factor independent of $f$.
\end{ThmRoots}

The theorem justifies the following exhaustive, albeit tedious, search
procedure for possible sets of coefficients $c(f)$, assuming they all have the
same magnitude, once a group $G$, a projective representation of $G$, and a
normalized factor system have been chosen.  Consider all possible sets of
coefficients of the form \eqref{eqn20}, setting $\gamma=1$ and
$c(e)=1/\sqrt{N}$ for the identity $e$ of $G$, as the global phase of $\UC$
is unimportant. For each set, check that the matrix $C$ given by \eqref{eqn13}
is unitary. Then see if the rows of the corresponding $\hat C'$, constructed
as described above, form a group under component-wise multiplication.
Using this procedure we have been able to show that if $N=2$, so the group is
$C_2$, the only two-qubit unitaries that can be implemented by our fast
protocol are either trivial products of unitaries or else equivalent under
local unitaries to a CNOT gate. (Note that there are additional fast two-qubit
unitaries that can be carried out using a bigger group, thus larger $N$ and
more entanglement.)

Both conditions (ii) and (iii) are nontrivial requirements.  Not every case in
which the $c(f)$ are of equal magnitude will lead to a unitary matrix $C$.
For example, if $c(f) = 1/\sqrt{N}$ for every $f$ and $\{\Gamma(f)\}$ is an
ordinary representation of $G$, so $\lambda(g,h)=1$, then \eqref{eqn13}
obviously does not define a unitary matrix.  And even if $C$ is unitary,
condition (iii) may not hold.  For example, the unitary in Eq.~(58) of
\cite{NLU} with $c(0,0)=c(1,0)=1/2,\, c(0,1)=e^{i\alpha}/2,\,
c(1,1)=-e^{i\alpha}/2$, assuming $\alpha$ is not an integer multiple of
$\pi/4$, results in a $\hat C'$ matrix whose rows do not form a group.

For every ordinary representation of an Abelian group $G$ there is a
corresponding fast protocol, as the group is a direct product (sum) of
cycles, and one can apply the construction in Example~6 below.
For general projective representations or non-Abelian groups the matter
remains open.

\subsection{Examples}
\label{sbct3.3}

The examples which follow represent just a few of the unitaries that can be
carried out by our double-unitary fast protocol. Examples 5 and 6 make
relatively efficient use of entanglement resources, in that the order of the
group, which is the rank of the fully-entangled resource state, is equal to
the Schmidt rank (or Schmidt number \cite{opSchmidt}) of $\UC$---the minimum
number of summands required to represent it as a sum of products of operators
on $A$ and $B$.  Examples 7 and 8, the latter involving a
non-Abelian group $G$, illustrate how the class of fast unitaries can be
significantly expanded by using more entanglement.

Note that any two-qubit unitary is equivalent under local
unitaries to one of the form (see ~\cite{Kraus01}),
\begin{equation}
  \UC = \exp[\, i(\alpha\sigma_x\ot\sigma_x + \beta\sigma_y\ot\sigma_y +
\gamma\sigma_z\ot\sigma_z)\, ],
\label{eqn21}
\end{equation}
where $\alpha$, $\beta$, and $\gamma$ are real numbers that can be calculated
from the matrix of $\UC$ (see, e.g., the appendix of \cite{Hammerer02} for the method of calculation).
For the two-qubit examples below we give the values of $\alpha$, $\beta$, and $\gamma$.\smallskip

{\bf Example 5.}

In the two-qubit unitary
\begin{equation}
\label{eqn22}
\UC = c(0) I\otimes I + c(1) X\otimes X + c(2) Z \otimes Z + c(3) XZ\otimes XZ,
\end{equation}
with $I$ the identity, $X$ and $Z$ the Pauli operators $\sigma_x$ and
$\sigma_z$, and $G$ the group $C_2\times C_2$, the method of search indicated
in Sec.~\ref{sbct3.2} yields the following possibilities for $c =
(c(0),c(1),c(2),c(3))$.

(a) The case $c = (1,1,1,-1)/2$ is equivalent to the SWAP gate defined in
\cite{Vidal02}, in which the two qubits are interchanged; $\alpha = \beta =
\gamma = \pi/4$ in \eqref{eqn21}. An alternative fast protocol for this gate
consists of teleportation done simultaneously in both directions.

(b) The case $c = (1,i,1,-i)/2$ implements the $U_{XY}$ gate as defined in
\cite{Vidal02}, equivalent under local unitaries to the double-CNOT (DCNOT)
gate defined in \cite{Hammerer02}; $\alpha = \beta =\pi/4,\, \gamma = 0$.

(c) The case $c = (1,1,\zeta,\zeta^5)/2$, where $\zeta=e^{i\pi/4}$;
$\alpha = \beta =\pi/4,\, \gamma = \pi/8$.

In each case the entanglement resource of two ebits required to carry out the
protocol is the minimum possible amount,
since the unitary is capable of creating two ebits of entanglement. \smallskip

{\bf Example 6.}

When the $\{\Gamma(f)\}$, with $f$ an integer between 0 and $N-1$, form an
ordinary representation of the cyclic group $C_N$ of order $N$, the
coefficients
\begin{equation}
  c(f) = \begin{cases}
 (1/\sqrt{N})\exp(-i \pi  f^2/N), & \text{ $N$ even,}\\
 (1/\sqrt{N})\exp(-i \pi  f(f+1)/N), & \text{ $N$ odd,}
 \end{cases}
\label{eqn23}
\end{equation}
 will provide a fast implementation of \eqref{eqn9}.  In particular
with $U(f) = V(f) = Z^{-f}$ this becomes
\begin{equation}
\label{eqn24}
\UC = \sum_{f=0}^{N-1} c(f) Z^{-f}\otimes Z^{-f}.
\end{equation}
The method of proof of Theorem~\ref{thm4} can be used to show the equivalence
of \eqref{eqn24} with $\UC=\sum_{k=0}^{N-1} \dyad{k}{k} \otimes Z^k$,
which in turn is locally equivalent to the $N$-dimensional CNOT gate of
Example~1.

{\bf Example 7.}

The unitary
\begin{align}\label{eqn25}
\UC = \frac{1}{2\sqrt{2}}(I\otimes I + X\otimes X + \zeta Z \otimes Z +
\zeta^5 XZ\otimes XZ & \notag\\
+ \zeta^3 I\otimes I + \zeta^7 X\otimes X +
\zeta^2 Z \otimes Z + \zeta^2 XZ\otimes XZ),&
\end{align}
of Schmidt rank 4 on two qubits, where $\zeta=e^{i\pi/4}$ and the operators
are the same as in Example~5, employs an unfaithful (each operator, e.g.
$X\otimes X$, appears twice in the sum) representation of the Abelian group
$C_2\times C_2\times C_2$, with the eight coefficients being the corresponding
$c(f)$. It can be verified that this $\{c(f)\}$ set satisfies the requirements for the fast
protocol. As this group is of order 8 the protocol requires a resource of 3
ebits, and we have not found any fast protocol which can implement this unitary using
less entanglement. It corresponds to $\alpha = \pi/4, \beta = \pi/8, \gamma=0$
in \eqref{eqn21} (the B gate of \cite{Zhang04}).\smallskip

{\bf Example 8.}

For any given integer $n\ge 2$, let $U(f) = V(f)$ ($0\le
f\le 2n-1$) be the $2\times 2$ matrices
\begin{align}
\label{eqn26}
\text{$0\le f\le n-1$: }& \mat{\cos(2f\pi/n) & -\sin(2f\pi/n)\\
\sin(2f\pi/n) & \cos(2f\pi/n)},\notag\\
\text{$n\le f\le 2n-1$: }&
\mat{-\cos(2f\pi/n) & -\sin(2f\pi/n)\\
-\sin(2f\pi/n) & \cos(2f\pi/n)}.
\end{align}
They form an irreducible ordinary representation of the dihedral group $D_n$
of order $N=2n$, where the first kind in \eqref{eqn26} correspond to rotations
and the second kind to reflections.  Let
\begin{equation}
\label{eqn27}
  c(f) = \begin{cases}
(\epsilon(f)/\sqrt{2n})\exp[\,i\pi  m f^2/n\,], & \text{ $n$ even,}\\
(\epsilon(f)/\sqrt{2n})\exp[\,i\pi  m f(f+1)/n\,], & \text{ $n$ odd,}
\end{cases}
\end{equation}
where $m$ is any positive integer coprime with $n$, and $\epsilon(f)$ is 1 for
$0\le f\le n-1$ and $i$ for $f \ge n$. It can be verified that these $\{c(f)\}$ sets satisfy the requirements for the fast
protocol. The two-qubit unitary constructed in this way is locally equivalent to \eqref{eqn21} with
$\alpha=\pi/4$, $\gamma=0$, and $\beta$, which necessarily lies in the
interval $[0,\pi/4]$, depending on $m$ and $n$ in a manner we have not studied
in detail.  It may be that the possible set of $\beta$ values is dense in $[0,\pi/4]$.

\subsection{Relationship to controlled-Abelian-group unitaries}\label{sbct3.4}

The following theorem, proved in Appendix~\ref{appCNOT}, shows that the family
of unitaries for which fast protocols were constructed in Sec.~\ref{sct2} can
also be realized using our fast protocol for double-group unitaries.  The
converse is not true, since, for instance, the 2-qubit SWAP gate in Example 5 cannot
be realized as a controlled-Abelian-group unitary, as it is of Schmidt rank 4,
while a controlled unitary on 2 qubits cannot have Schmidt rank greater than 2.

\begin{ThmControl}\label{thm4}
  Let $\UC$ be a controlled-Abelian-group unitary of the form \eqref{eqn1},
  where the $V_k$ are a subset of an ordinary representation of an Abelian
  group $G$ of order $N$. Then $\UC$ is equivalent under local unitaries to
\begin{equation}\label{eqn28}
\WC = \sum_{f=0}^{N-1} c(f) Q(f)\otimes R(f),
\end{equation}
where the $c(f)$ are complex coefficients, the $Q(f)$ are linear combinations
of $P_k$'s, and $\{Q(f)\}$, $\{R(f)\}$, $\{Q(f)\otimes
R(f)\}$ are all ordinary representations of the group $G$.  In addition the $c(f)$
can be chosen to satisfy the requirements for the fast protocol as given in
Sec.~\ref{sbct3.2}. Hence all controlled-Abelian-group unitaries of the form
discussed in Sec.~\ref{sct2} can be implemented by our fast double-group
unitary protocol, without using more entanglement.
\end{ThmControl}

\section{Conclusions}\label{sct4}

Any nonlocal unitary can be carried out deterministically by means of local
operations and classical communication provided an appropriate entangled
resource is available.  However, teleportation and various more efficient
schemes typically require two rounds of classical communication, and hence the
minimum total amount of time required to complete the protocol is twice the
time required for one-way communication.  In certain cases there are fast
protocols in which the minimum total time is only half as long, and in this
paper we have discussed two protocols for fast bipartite unitaries. The first
is shown in Fig.~\ref{fgr2}: it carries out a controlled-Abelian-group unitary
of the form \eqref{eqn1}, including cases in which only a subset of the
collection $\{V_k\}$ that forms an Abelian group appear in the sum.
The second, shown in Fig.~\ref{fgr4}, will carry out a double-group unitary of
the form \eqref{eqn9}, provided the coefficients $c(f)$ satisfy appropriate
conditions.  We have shown, Sec.~\ref{sbct3.4}, that unitaries which can be
carried out by the first protocol can also be carried out by the second,
though the converse is not true (e.g., Example 5).  We have constructed some
examples for both protocols.

Note, however, that we have not been able to answer the fundamental question
as to precisely \emph{which} unitaries can be carried out \emph{exactly} using
a fast protocol and a fixed entanglement resource specified in advance. We do
not know the answer even for fast protocols of the two types considered in
this paper. Finding examples for our double-group protocol is not at all
trivial; see the discussion in Sec.~\ref{sbct3.2}.  In Sec.~\ref{sbct2.3} we
discussed cases in which subsets of a group can be used to carry out a fast
controlled unitary protocol at the cost of greater entanglement.  See in
particular Example 3, where we showed that any unitary in a particular
continuous family can be approximated arbitrarily closely by a unitary
implementable by a deterministic fast protocol, provided one is willing to use
up enough entanglement.  This is similar in spirit to the results in
\cite{Buhrman} and \cite{Beigi11}.  Their protocols may need less or more entanglement
than our protocols, depending on the form of the unitary. In certain situations (e.g.,
Example 5), our protocol uses the minimum possible entanglement.
It would be nice if these issues could be clarified in terms of some basic principle(s) of quantum
information theory.

Another question we have not been able to answer is whether unitaries of the
more general form $\sum_f U(f)\ot W(f)$, where the $U(f)$ form an ordinary or
projective representation of a group, but $W(f)$ need not do so, can be
carried out by means of a fast protocol.  A slow protocol was found in our
earlier work \cite{NLU}, and we have constructed a fast version for that
protocol, but it seems to only work for those unitaries implementable by our
fast double-group protocol of Sec.~\ref{sct3}.  Again, this may reflect some
fundamental principle of quantum information, but if so we have not been able
to identify it.

\section{Acknowledgments}

We thank Serge Fehr, Hoi Kwan Lau, and Shiang Yong Looi for helpful
discussions, including those on the connections of this work with the topics
of instantaneous measurement and position-based quantum cryptography. Patrick
Coles read some of the appendices and made helpful suggestions. This work has
been supported in part by the National Science Foundation through Grants
No.~PHY-0456951 and No.~PHY-0757251. S.M.C. has also been supported by a grant from the
Research Corporation.

\begin{appendix}

\section{General considerations in protocols}\label{sct_extend}

\begin{figure*}[ht]
\begin{centering}
\includegraphics[scale=1]{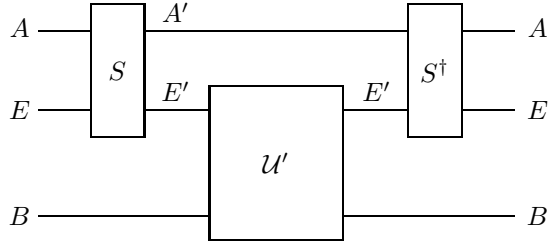}
\caption{A scheme for implementing $\UC$ using a protocol for $\UC'$.}\label{figextend}
\end{centering}
\end{figure*}

In this section, we consider implementation of $\UC$ on $\HC_A\otimes\HC_B$ using a different $\UC'$ and use these ideas below to show that consideration of controlled unitaries can be restricted to those with rank-$1$ projectors. A scheme is shown in Fig.~\ref{figextend}, which is valid for both the fast and slow protocols. If the protocol for $\UC'$ is fast, then the whole protocol is fast.

The circuit in Fig.~\ref{figextend} can be used for the following two situations, to be discussed in detail in the two subsections below. The first situation, called ``extension'', is that we extend the space of $\HC_A$ to $\HC_{E'}$, and the unitary $\UC':\HC_{E'}\otimes\HC_B\rightarrow \HC_{E'}\otimes\HC_B$ is an extension of $\UC$, where $\UC$ is any unitary on $\HC_A\otimes\HC_B$. The second situation, called ``compression with extension,'' is only for general controlled unitaries $\UC$ of the form \eqref{eqn1}. The protocol replaces the higher-rank projectors on $\HC_A$ in $\UC$ with rank-one projectors on $\HC_{E'}$ in $\UC'$, while adding more projectors if needed. Applications are found in Sec.~\ref{sct2}. Note that while Fig.~\ref{figextend} shows an extension (and compression in the case of controlled unitaries) on the $A$ side, one can just as well do this on the $B$ side, or both sides.

The input state for the whole protocol is any state on $AB$ together with some fixed state on the ancilla $E$ denoted by $\ket{0}_E$. The map $S:\HC_A\otimes\HC_E\rightarrow \HC_{A'}\otimes\HC_{E'}$ is unitary, and  $S^\dag$ is its inverse. The unitary $S$ obviously has its input dimension equal to its output dimension: $d_A d_E=d_{A'} d_{E'}$, where $d_{E'}$ is determined by $\UC'$ (see below), $d_A$ may be unequal to $d_{A'}$, and $d_E$ may be unequal to $d_{E'}$.

\subsection{Extending the Hilbert space in protocols}\label{sbct_extend1}

Here we consider the first type of extension, where $\UC'$ is the direct sum of $\UC_0$ (the same as $\UC$ but on a different space) and another unitary $\RC$. Dimension $d_{E'}$ is fixed by $\UC'$ and is greater than $d_A$. One can always choose $d_E$ to be equal to $d_{E'}$, but in general may choose $d_E$ to be less than $d_{E'}$. The action of the unitary $S$ on the actual input space is determined by the following equation:
\begin{equation}
\label{eqn29}
 S(\ket{k}_A \otimes \ket{0}_E)=\ket{0}_{A'}\otimes \ket{k}_{E'},\,\,\,k=0,1,\cdots,d_A-1,
\end{equation}
where $\{\ket{k}_A\}$ is an orthonormal basis of $\HC_A$. The requirements for $S$ in this equation can be extended to a full definition of a unitary. The effect of $S$ is to transfer Alice's input state into $\HC_{E'}$. Define $\HC_{\bar A}$ to be the span of $\{\ket{k}_{E'}:0\le k\le d_A-1\}$. Then $\HC_{E'}=\HC_{\bar A} \oplus\HC_R$, where $\HC_R$ is a space orthogonal to $\HC_{\bar A}$. Then the correct form of $\UC'$ should be $\UC'=\UC_0\oplus\RC$, where $\UC_0:\HC_{\bar A}\otimes \HC_B\rightarrow\HC_{\bar A}\otimes\HC_B$ is the same as the original unitary $\UC$ except it is on a different space, and $\RC:\HC_R\otimes \HC_B\rightarrow\HC_R\otimes\HC_B$ is an arbitrary unitary.

Now we prove that the circuit defined in Fig.~\ref{figextend} applied to $\ket{0}_E\otimes\ket{\psi}_{AB}$ yields $\ket{0}_E\otimes \UC\ket{\psi}_{AB}$.
\begin{proof}
Suppose the input state on $\HC_{AB}$ is $\ket{\psi}_{AB}=\ket{k}_A \ket{q}_B$. Then
 \begin{align}
\ket{k}_A\ket{0}_E\ket{q}_B &\stackrel{S}\longrightarrow \ket{0}_{A'}\ket{k}_{E'}\ket{q}_B\notag\\
&\stackrel{\UC'}\longrightarrow \ket{0}_{A'}\otimes \UC'(\ket{k}_{E'}\ket{q}_B)\notag\\
&=\ket{0}_{A'}\otimes \UC(\ket{k}_{E'}\ket{q}_B)\notag\\
&=\ket{0}_{A'}\otimes \sum_{j=0}^{d_A-1}\sum_{p=0}^{d_B-1} \langle jp\vert\UC\vert kq\rangle \ket{j}_{E'}\ket{p}_B \notag\\
&\stackrel{S^\dag}\longrightarrow \ket{0}_{E}\otimes \sum_{j=0}^{d_A-1}\sum_{p=0}^{d_B-1} \langle jp\vert\UC\vert kq\rangle \ket{j}_{A}\ket{p}_B \notag\\
&=\ket{0}_E\otimes \UC(\ket{k}_A\ket{q}_B)
\label{eqn30}
\end{align}
The argument can be extended by linearity to superpositions of the states $\ket{k}_A \ket{q}_B$.
\end{proof}

As a side remark, the derivations above should still work if we replace the unitary $S$ by an isometry $\VC = S\ket{0}_E$, replace $S^\dag$ by $\VC^\dag$, and remove system $E$ from the circuit in Fig.~\ref{figextend}. It can be verified that the overall operation of the circuit is $(\VC^\dag \otimes I_B)(I_{A'}\otimes \UC')(\VC\otimes I_B)=\UC$. We chose to present the argument using the unitary $S$ rather than the isometry $\VC$ in order to show that the scheme has no trouble finding an experimental implementation. The same remark also applies to the next subsection.

The current extension technique was useful in finding the protocols in Secs.~\ref{sbct2.3} and Example 8 in \ref{sbct3.3}, but it turns out that those protocols (for the particular types of unitaries) can be simplified such that no extension is needed, which is why we have not explicitly mentioned this idea of extension in those sections.

\subsection{Controlled unitaries: Conversion of higher rank projectors into rank-one projectors}\label{sbct_control_rank}

For general controlled unitaries $\UC$ of the form \eqref{eqn1} (not limited to those implementable by the fast protocols in this paper), we now consider a procedure that compresses the higher-rank projectors on $\HC_A$ into rank-one projectors on $\HC_{E'}$, while adding more projectors if needed. The form of $\UC'$ is
\begin{equation}\label{eqn31}
\UC'=\sum_{k=0}^{N'-1} \dyad{k}{k}_{E'} \otimes V_k
\end{equation}
where $N'\ge N$. Apparently $d_{E'}=N'$.

The steps of the protocol are similar to those in Appendix~\ref{sbct_extend1}, but with the following change to the requirements on $S$ (and accordingly $S^\dag$):
\begin{equation}\label{eqn32}
S(\ket{k,r}_A \otimes \ket{0}_E)=\ket{r}_{A'}\otimes \ket{k}_{E'},
\end{equation}
where $(k,r)$ is the label for the states in a specific basis of $\HC_A$, with $k$ ($0\le k\le N-1$) labeling which projector $P_k$, and $r\in \{0,1,\cdots,\mbox{rank}(P_k)-1\}$ labeling which basis state in the support of $P_k$. Note the range of $r$ depends on $k$, and because of this, $d_{A'}$ should be at least the maximum rank among the $P_k$'s ($0\le k\le N-1$), while satisfying $d_A d_E=d_{A'} d_{E'}$.
The requirements for $S$ in \eqref{eqn32} can be extended to a full definition of a unitary. The effect of $S$ is to transfer the information about ``which $k$'' into $\HC_{E'}$, and that information is used in the controlled unitary $\UC'$, and then transferred back to $\HC_A$ by $S^\dag$.

The final state of the protocol is $\ket{0}_E\otimes\UC\ket{\psi}_{AB}$, and the proof for the correctness of the protocol is similar to that in Appendix~\ref{sbct_extend1}.

\begin{proof}
Suppose the input state on $\HC_{AB}$ is $\ket{\psi}_{AB}=\ket{k,r}_A \ket{q}_B$. Then
 \begin{align}
\ket{k,r}_A\ket{0}_E\ket{q}_B &\stackrel{S}\longrightarrow \ket{r}_{A'}\ket{k}_{E'}\ket{q}_B\notag\\
&\stackrel{\UC'}\longrightarrow \ket{r}_{A'}\otimes \UC'(\ket{k}_{E'}\ket{q}_B)\notag\\
&=\ket{r}_{A'}\otimes \ket{k}_{E'}\otimes V_k\ket{q}_B)\notag\\
&\stackrel{S^\dag}\longrightarrow \ket{0}_{E}\otimes \ket{k,r}_{A}\otimes V_k\ket{q}_B \notag\\
&=\ket{0}_E\otimes \UC(\ket{k,r}_A\ket{q}_B)
\label{eqn33}
\end{align}
The argument can be extended by linearity to superpositions of the states $\ket{k,r}_A \ket{q}_B$.
\end{proof}

As noted above, for the current subsection it is also plausible to replace the unitary $S$ by an isometry $\VC = S\ket{0}_E=\sum_k \sum_r \ket{r}_{A'}\ket{k}_{E'}\bra{k,r}_A$, replace $S^\dag$ by $\VC^\dag$, and remove system $E$ from the circuit in Fig.~\ref{figextend}. Using the definition of $\UC'$ in \eqref{eqn31}, it can be verified that the overall operation of the circuit is $(\VC^\dag \otimes I_B)(I_{A'}\otimes \UC')(\VC\otimes I_B)=\UC$, where $\UC$ is of the form \eqref{eqn1}.

\section{Proofs of Theorems~\ref{thm1}, \ref{thm2}, \ref{thm3} }
\label{appthm1}

\textbf{Proof of Theorem~\ref{thm1}. }

(a) We need to show [see \eqref{eqn14} and note that both $D$ and $Z_l$ are
diagonal matrices] that
\begin{align}
\label{eqn34}
\tilde P_l =
 C Z_l C\ad &= (1/N) M\hat K D Z_l D\ad \hat K\ad M\ad \notag\\
 &= (1/N) M\hat K Z_l  \hat K\ad M\ad
\end{align}
is a complex permutation matrix.  The diagonal elements of $Z_l$ are complex
conjugates of the elements in a row of $\hat T$ and thus [Eq.~\eqref{eqn15}]
equal to a common phase factor times those in a particular row, say row
$m$, of the character table $\hat K$; recall that the complex conjugate of
a row in $\hat K$ is always another row of $\hat K$.  Now the matrix $\hat
K Z_l$ is the matrix $\hat K$ with each column multiplied by the corresponding
diagonal element of $Z_l$.  Thus the $j$-th row of $\hat K Z_l$ is the
element-wise product of row $j$ of $\hat K$ with row $m$ of $\hat K$, up to an
overall phase that depends on $j$. But since the rows of $\hat K$ form a group under
element-wise products, this means that $\hat K Z_l = Q_l \hat K$
for a suitable complex permutation matrix $Q_l$. Since $\hat K \hat K\ad = N
I$, \eqref{eqn34} tells us that $\tilde P_l = M Q_l M\ad$, and because $M$,
$Q_l$, and $M\ad$ are all complex permutation matrices, so is $\tilde
P_l$.\smallskip

(b) The rows of $\hat T$ are the elements of a group $H$ in the following
sense. The element-wise product of any two rows is, up to a phase, a third
row, and the fact that the rows are linearly independent means that this third
row is uniquely determined.  That is, there is a well-defined associative
group multiplication, which is commutative, so the group $H$ is Abelian.
There is necessarily one row consisting of identical elements; this is the
identity element of $H$.  Given any row, there is another row which is,
element by element, its complex conjugate, up to a single phase for the whole
row; these two rows are inverses of each other.  Hence the group $H$ is
well defined.  Obviously, each column of $\hat T$ consists of elements (viewed
as $1\times 1$ matrices) that form an irreducible representation of $H$ under
the multiplication of complex numbers, and all the elements of $\hat T$ are of
magnitude 1.

 Divide each row by its first element to form the matrix $\hat T'$, whose rows again form the group $H$, but now without additional phases since the first element of each row is 1. Since the rows of $\hat T$ are linearly independent, so are
the rows of $\hat T'$, and hence also its columns. Thus each column of $\hat T'$ forms a distinct irreducible representation of
the group $H$. All the $N$ distinct irreducible representations of $H$ are included in the columns of $\hat T'$, thus $\hat T'$ is the transpose of a character table of the Abelian group $H$, and such a transpose is itself a character table of $H$ (see the discussion preceding Theorem~\ref{thm1}). Hence we can identify $\hat T'$ as the $\hat K$ in \eqref{eqn15}, and the phases used to change $\hat T$ to $\hat T'$ can be included in the matrix $L$ in \eqref{eqn15}.\hfill\qedsymbol
\smallskip

\textbf{Proof of Theorem~\ref{thm2}. }

Let $\hat C=\sqrt{N} C$ and rewrite \eqref{eqn14} in the equivalent form
\begin{equation}
\label{eqn35}
\hat C Z_l= {\tilde P}_l \hat C.
\end{equation}
The matrix on the left is obtained from $\hat C$ by multiplying each column by
the corresponding diagonal element of $Z_l$, while the one on the right is
obtained by some permutation of the rows of $\hat C$, with an additional
overall phase for each row.  Consider the special case in which the first row
of $\hat C$ consists of 1's. Then the first row of $\hat C Z_l$ is $\breve
Z_l$, the row vector whose elements are the diagonal elements of $Z_l$, and
according to \eqref{eqn35}, it is equal to the first row of ${\tilde P}_l \hat
C$ (i.e., a phase times some other row of $\hat C$).  As the $\breve Z_l$ are
linearly independent (they are complex conjugates of the rows of the Hadamard
matrix $\hat T$), it follows by equating, for each $l$, the first row on the
left side of \eqref{eqn35} with that on the right side, that the element-wise
product of the first row of $\hat C$ and the $\breve Z_l$ for all possible
values of $l$ generates all the rows of $\hat C$ up to a phase (i.e., each row
of $\hat C$ is a phase times one of the $\breve Z_l$). Then according to
\eqref{eqn35}, the element-wise product of any row of $\hat C$ with any
$\breve Z_l$ (i.e. the element-wise product of any two rows of $\hat C$ up to
a phase), is always a third row of $\hat C$ up to a phase. Similarly the
product of any two of the matrices in the set $\{Z_l\}$ is a third, up to a
phase; this is an associative group product.  The first row of $\hat C$ is
equal to one of the $\breve Z_l$ up to a phase, and this $Z_l$ corresponds to
the group identity.  Because the first row of $\hat C$ consists of 1's, there
is a row of $\tilde P_l \hat C$, say row $m$, which consists of equal
elements.  Then the $m$-th row of $\hat C $ [see \eqref{eqn35}], identifies the
group inverse of $Z_l$.  Thus the $Z_l$ under matrix products form a group
(denoted by $H$) up to phases. For any $Z_l$, its complex conjugate $Z_l^\ast$
is the matrix inverse of $Z_l$, and hence some $Z_{l'}$ up to a phase.
Consequently all the different rows of $\hat T$, the complex conjugates of
$\breve Z_l$, are equal to the different $\breve Z_{l'}$ up to phases and a
permutation of the ordering, hence these rows form the group $H$ up to phases
under element-wise multiplication.  Then according to Theorem~\ref{thm1}(b),
$H$ must be Abelian, and $\hat T$ is of the form given in \eqref{eqn15}.
Since the rows of $\hat T$ are a permutation of the different $\breve Z_{l'}$
up to phases, they are also a permutation of the rows of $\hat C$ up to
phases.  Thus $\hat C$ is of the form \eqref{eqn15} with $D=I$.  Hence for the
special case under consideration, \eqref{eqn15} is satisfied, and then
\eqref{eqn16} follows from \eqref{eqn15}.

Next consider the more general case in which the elements of the first row of
$\hat C$ are all nonzero.  Form $\hat C'$ from $\hat C$ by dividing every
column of $\hat C$ by the corresponding element on the first row.  This means
that $\hat C' = \hat C Q$, where $Q$ is a diagonal matrix.  Since $Q$ commutes
with every $Z_l$, we can replace $\hat C$ on both sides of \eqref{eqn35} with
$\hat C'$.  As the first row of $\hat C'$ consists of 1's, the argument given
above shows that the $\{Z_l\}$ form an Abelian group up to phases, and the
rows of $\hat C'$ are, up to phases, some permutation of the rows of $\hat T$, and
$\hat T$ is of the form given in \eqref{eqn15}, again according to Theorem~\ref{thm1}(b).
Thus the different columns of $\hat C'$ all have the same normalization, and
since the same is true of $\hat C$, as $C$ is assumed to be a unitary matrix,
it follows that the diagonal elements of $Q$ all have magnitude 1, and one can
set $D=Q^{-1}$ in \eqref{eqn15}. Then \eqref{eqn16} follows from \eqref{eqn15}.\hfill\qedsymbol
\smallskip

\textbf{Proof of Theorem~\ref{thm3}. }

The coefficients $c(f)$ must all be of the same magnitude, $1/\sqrt{N}$ (see
the remark following Theorem~\ref{thm1}), and without loss of generality
$\gamma$ can be chosen such that $c(e)=1/\sqrt{N}$ in \eqref{eqn20}, where $e$
is the group identity.  First consider the case that $\{\Gamma(f)\}$ is an
ordinary representation of $G$, so that $\lambda(g,h)=1$.  Choose any group
element $r\ne e$ and assume it has order $p$, which means that $p$ divides $N$
and $r^p=e$.  The first row of $C$, corresponding to $g=e$ in \eqref{eqn13},
contains $c(e),c(r),c(r^2),\cdots,c(r^{p-1})$ in some order, interspersed with
other coefficients $c(f)$. Now consider the row of $C$ corresponding to
$g=r^{p-1}=r^{-1}$ in \eqref{eqn13}.  It is related to the first row as
indicated here,
\begin{equation}
\label{eqn36} \left( \begin{array}{cccccc} c(e) & c(r) &
c(r^2) & \cdots & c(r^{p-1}) & \cdots \\ c(r) & c(r^2) & \cdots & c(r^{p-1}) &
c(e) & \cdots
\end{array} \right),
\end{equation}
where only the relevant columns are shown, rearranged in a convenient order.

Let us define the $\hat C'$ matrix, as in the first paragraph of
Sec.~\ref{sbct3.2}, to be the one obtained from $\hat C=\sqrt{N}\,C$ by
multiplying each row and each column by some phase, so that all the elements
in the first row and first column are equal to 1. This is a character table,
so every element is an $N$-th root of 1.  Equivalently, $\hat C'$ is obtained
from $C$ by dividing each column of $C$ by the corresponding element in the
first row, and then in the resulting matrix dividing each row by its first
element. Consequently, applying this process to the rows and columns shown in
\eqref{eqn36}, we conclude that
\begin{align}
\label{eqn37}
\frac{c(r^2)}{c(r)^2} =  \phi_1\sqrt{N},\quad
\frac{c(r^3)}{c(r)c(r^2)} =  \phi_2\sqrt{N},\notag\\
\quad\ldots,\quad
\frac{c(e)}{c(r)c(r^{p-1})} =  \phi_{p-1}\sqrt{N},
\end{align}
where each $\phi_j$ is an $N$-th root of 1.
The product of these $p-1$ equations,
\begin{equation}
\label{eqn38}
c(e)/[c(r)]^p=\phi_1 \phi_2 \cdots \phi_{p-1}(\sqrt{N})^{p-1},
\end{equation}
implies, since $c(e)=1/\sqrt{N}$, that $\sqrt{N}\,c(r)$ must be a $p$-th root
of a number which is itself an $N$-th root of 1, and because $p$ divides $N$,
$c(r)$ is of the form \eqref{eqn20}.  This completes the argument for an
ordinary representation.

When the $\{\Gamma(r)\}$ form a projective representation of $G$ with a
standard factor system \eqref{eqn11}, the first row in \eqref{eqn36} is the
same, but the second row will be multiplied by appropriate factors
$\lambda(g,h)$. Since we are assuming a normalized factor system, all these
additional factors are themselves $N$-th roots of 1, so \eqref{eqn37} still
holds for $\phi_j$ which are $N$-th roots of 1, and the rest of the argument
is the same as before.\hfill\qedsymbol

\section{Proof of Theorem~\ref{thm4} }\label{appCNOT}

In this appendix, we prove Theorem~\ref{thm4}, which says that the controlled unitary $\UC=\sum_{k=0}^{N-1} P^A_k \otimes V^B_k$ given by \eqref{eqn1}, where $P^A_k$ are orthogonal projectors, and $\{V^B_k\}$ form a subset of an ordinary representation of an Abelian group $G$, is equivalent to
\begin{equation}\label{eqn39}
\WC = \sum_{f=0}^{N-1} c(f) Q(f)\otimes R(f)
\end{equation}
under local unitaries, where $c(f)$ are complex coefficients [will be defined in \eqref{eqn41}], and $Q(f)$ are linear combinations of $P^A_k$, and $\{Q(f)\otimes R(f)\}$ is an ordinary representation of $G$. In addition, the $c(f)$ can be chosen to satisfy the requirements for the fast protocol, hence all controlled-Abelian-group unitaries can be implemented by the fast double-group unitary protocol.\\

We first prove the case that $\{V^B_k\}$ form a whole representation, not a subset, and at the end we will remark that the proof also works in the ``subset'' case. The proof is by explicitly constructing a $\WC$ and showing that it is equivalent to $\UC$ under local unitaries. Any Abelian group is a direct sum of cyclic groups, so $G=C_{r_1}\oplus C_{r_2} \oplus \cdots \oplus C_{r_{\eta}}$, where ${\eta}\ge 1$, and $r_i$ is the order of the cyclic group $C_{r_i}$. Then $N=\vert G\vert=\prod_{i=1}^{\eta} r_i$. The group element $k$ is relabeled by a vector $k=(k_1,k_2,\cdots,k_{\eta})$, where $0\le k_i\le r_i-1$. We use the convention that $k$ is the sequential numbering (starting from 0) for the vectors in lexicographical order, so that $k=0$ corresponds to $(0,0,\cdots,0)$, and $k=1$ corresponds to $(0,0,\cdots,1)$, etc.  Suppose $\{V^B_k\}$ has been diagonalized under a suitable unitary similarity transform, then $\{V^B_k\}$ is the direct sum of some irreducible representations (possibly with redundancy).
All possible irreducible representations of $G$ are one-dimensional, and have the form
\begin{equation}\label{eqn40}
R^q(k)=\prod_{s=1}^{\eta} \exp(2\pi i q_s k_s/r_s),
\end{equation}
where $q=(q_1,q_2,\cdots,q_{\eta})$ is the label for irreducible representations (some may be missing from $\{V^B_k\}$, but we still include them in this labeling scheme for convenience). Denote the computational basis of $\HC_B$ by $\{\ket{b}\}$, $b=0,1,\cdots,d_B-1$, then the $b$-th diagonal elements in $V^B_k$ determine an irreducible representation labeled by $q_b$, $0\le q_b\le N-1$. The $q_b$ can be written in the vector form [see \eqref{eqn40} and the sentence after that], and the components in the vector $q_b$ will be denoted by $q_{b,s}$.

As discussed above, for every $f\in G$ we can represent $f$ using a set of integers: $f=(f_1,f_2,\cdots,f_{\eta})$, with group multiplication corresponding to vector addition (modulo $r_s$ for the $s$-th element of the vector). Define $c(f)=\prod_{s=1}^{\eta} c_s(f_s)$, where $c_s(f_s)$ is defined by (basically the same as in Example 6)
\begin{align}\label{eqn41}
c_s(f_s) = \frac{1}{\sqrt{r_s}}\exp[-\pi i f_s(r_s\mbox{ mod }2+f_s)/r_s], \notag\\ \,\,\,f_s=0,1,\cdots,r_s-1.
\end{align}

We choose the $Q(f)$ to be
\begin{equation}\label{eqn42}
Q(f)=\frac{1}{\sqrt{N}} \sum_{k=0}^{N-1} \left[\prod_{s=1}^{\eta} \exp(-2\pi i f_s k_s /r_s)\right] P^A_k,
\end{equation}
where $(k_s)$ is the vector labeling for $k$. Define $R(f)$ as
\begin{equation}\label{eqn43}
R(f)=\frac{1}{\sqrt{N}} \sum_{b=0}^{d_B-1} \left[\prod_{s=1}^{\eta} \exp(-2\pi i f_s q_{b,s} /r_s)\right] \dyad{b}{b},
\end{equation}
where $q_{b,s}$ are the components in the vector labeling for $q_b$. It is not hard to verify that $\{Q(f)\otimes R(f)\}$ is an ordinary representation of $G$, and the coefficients $c(f)$ form a unitary $C$ matrix of the type \eqref{eqn13}, hence $\WC$ is unitary.

With the above choices of $Q(f)$ and $R(f)$,
\begin{widetext}
\begin{equation}\label{eqn44}
\WC\ket{k}\ket{b}=\frac{1}{\sqrt{N}}\sum_{f=0}^{N-1}
\left(\prod_{s=1}^{\eta}\exp[-\pi i f_s(r_s\mbox{ mod }2+f_s+2k_s+2q_{b,s})/r_s]\right)\ket{k}\ket{b},
\end{equation}
where $0\le k\le N-1$, $\,\, 0\le b\le d_B-1$, and $\ket{k}$ is any eigenstate of $P^A_k$.
Denote the phase factor in front of $\ket{k}\ket{b}$ in the above equation by $\zeta_{kb}$, then
\begin{align}\label{eqn45}
\zeta_{kb} & = \frac{1}{\sqrt{N}}\prod_{s=1}^{\eta}\sum_{j=0}^{r_s-1}\exp \Big( \pi i \{-[j+k_s+q_{b,s}+(r_s\mbox{ mod }2)/2]^2+[k_s+q_{b,s}+(r_s\mbox{ mod }2)/2]^2\}/r_s \Big) \notag\\
& = \frac{1}{\sqrt{N}}\prod_{s=1}^{\eta}\left[\left(\sum_{j=0}^{r_s-1}\exp\{-\pi i [j+k_s+q_{b,s}+(r_s\mbox{ mod }2)/2]^2/r_s\}\right) \exp\{\pi i [k_s+q_{b,s}+(r_s\mbox{ mod }2)/2]^2/r_s\}\right]\notag\\
& = \frac{\alpha}{\sqrt{N}}\prod_{s=1}^{\eta}\exp\{\pi i[k_s+q_{b,s}+(r_s\mbox{ mod }2)/2]^2/r_s\}
\end{align}
\end{widetext}
where $\alpha$ is a constant independent of $k$ and $q$. In deriving the last line above, we have used $(r+j)^2\equiv j^2(\mbox{mod }2r)$ for even $r$, and $(r+j+1/2)^2\equiv (j+1/2)^2(\mbox{mod }2r)$ for odd $r$, which make the substitution $j+k_s+q_{b,s}\rightarrow j$ possible, and obtained $\alpha=\prod_{s=1}^{\eta} \alpha_s$, where $\alpha_s=\sum_{j=0}^{r_s-1} \exp\{-\pi i [j+(r_s\mbox{ mod }2)/2]^2/r_s\}$.

Define the local operators $M_A$ and $M_B$ on $\HC_A$ and $\HC_B$, respectively, as follows:
\begin{align}\label{eqn46}
M_A=\sum_{k=0}^{N-1}\zeta_{k0}^{-1}P^A_k,\quad \notag\\ M_B=\zeta_{00}\sum_{b=0}^{d_B-1}\zeta_{0b}^{-1}\dyad{b}{b}.
\end{align}
From the unitarity of $\WC$, $\zeta_{kb}$ is always a phase factor with magnitude 1, hence $M_A$ and $M_B$ are unitary operators.
Then for $\ket{k}$ chosen arbitrarily from the eigenstates of $P^A_k$, we have
\begin{align}
(M_A\otimes M_B)\WC\ket{k}\ket{b} & = (M_A\otimes M_B) \zeta_{kb}\ket{k}\ket{b} \notag\\
& = \zeta_{00}\zeta_{kb}\zeta_{k0}^{-1}\zeta_{0b}^{-1}\ket{k}\ket{b} \notag\\
& = \left[ \prod_{s=1}^{\eta} \exp(2\pi i k_s q_{b,s}/r_s) \right] \ket{k}\ket{b} \notag\\
& = \sum_{k=0}^{N-1}P^A_k\otimes V^B_k \ket{k}\ket{b} \notag\\
& = \UC \ket{k}\ket{b}
\end{align}
where we have used \eqref{eqn45} to derive the third line, and used \eqref{eqn40} to derive the fourth line. Since $P^A_k$ are of finite rank, there exists a finite collection of states of the form $\ket{k}\ket{b}$ to make a complete basis of $\HC_{AB}$. The actions of $(M_A\otimes M_B)\WC$ and $\UC$ are the same on all states in a complete basis, hence they must be identical operators. Therefore $\UC$ is equivalent to $\WC$ under local unitaries.

Using the algorithm in Sec.~\ref{sbct3.2}, it can be verified that the choice of coefficients $c(f)$ given above (which can be viewed as the choice in Example 6 generalized to the non-cyclic Abelian groups) satisfies the requirements for the fast protocol. Hence the double-group protocol for $\WC$ is fast.

The proof above can basically be applied to the case that $\{V^B_k\}$ form a subset of a representation. In general some $P^A_k$ do not occur in the expressions for $Q(f)$ and $M_A$; those $P^A_k$ can be safely removed because $Q(f)$ and $M_A$ are block diagonal, where the blocks are determined from the support of the $P^A_k$'s. The coefficients $c(f)$ are still the same as above, so the protocol is still fast. Hence the proof still works. \hfill\qedsymbol

\end{appendix}

\end{document}